\documentstyle[aps,epsfig]{revtex}

\date{\today}

\title{SC state
in the underdoped 
  high-$T_c$ cuprates as a quantum spin liquid. A microscopic
theory.}

\author{F Onufrieva and P Pfeuty}
\address{Laboratoire Leon Brillouin CE-Saclay 91191 Gif-sur-Yvette France}

\begin{document}

\twocolumn[\hsize\textwidth\columnwidth\hsize\csname@twocolumnfalse\endcsname
\maketitle

\begin{abstract}
We have discovered a new example of  quantum spin liquid
which is a superconducting (SC) phase in 2D electron system close to
electronic topological transition.
 As a  quantum spin liquid in
low dimensional localized spin systems
  it is characterized by a resonance
spin mode and a well separated two-particle
continuum.
Application of the theory to the  high-$T_c$ cuprates
allows to shed light on many observed 
in the underdoped regime features such 
as the so called  $40 meV$ resonance peak, incommensurability at lower $\omega$,
spin gap etc. 
\end{abstract}

]

Two most intriguing problems of  high-$T_c$ cuprates,
{\bf normal state} anomalies  including a pseudogap phenomenon
observed by  ARPES \cite{ARPES} and by many other experiments
and a peculiar {\bf SC} state spin dynamics
observed by neutron scattering (including the
so called   resonance peak 
\cite{6.92,6.7,6.75,6.8,6.83,incommensurability,BSCO}, incommensurability at lower $\omega$ \cite{incommensurability,Arai},
spin gap etc.),
are considered today
as key problems for understanding the physics of the cuprates.
 Have these phenomena (both occuring
in the underdoped
regime) some
interelation or are they  completely independent? Is there
some
interrelation with the third phenomenon,
existence of  high-$T_c$ itself?
There is a number of theories devoted to each of three
 problems.
Most of them are   phenomenological, use certain anzatzs which
moreover  are different to explain the
different phenomena.

 In the present paper where we concentrate on the problem
of  spin dynamics in SC state we show that the observed
by inelastic neutron scattering (INS) peculiar behaviour
 can be naturally understood within the 
 same theoretical concept which we used before \cite{PRL1,PRL2} to explain 
normal state anomalies and high value of SC gap. This concept is:
 (i)  proximity of the
 2D  system of fermionic quasiparticles on a square
lattice to electronic topological transition (ETT) (while
this ETT is quite unusual
in the case of electron-hole asymmetry or by other
words of hoping beyond nearest neighbors)
 and (ii) existence of
strong exchange AF interaction  between these quasiparticles.
[For the high-$T_c$ cuprates   
   the latter are not the bare electrons but the quasiparticles 
\cite{Onufrieva96} in the strongly correlated 
$CuO_2$ plane (described by the
$t-t'-J$ model) 
which dispersion law is also determined by the topology
of 2D square lattice.]
 In \cite{PRL1} we have shown that
obligatory consequences of (i)  and (ii) are  developing of d-wave
superconductivity around ETT point $\delta=\delta_c$, $T=0$, 
with maximum $T_{sc}$
at  $\delta=\delta_c$ and the existence 
in the underdoped
regime above $T_{sc}$
of anomalous metallic state  with strong spin density wave
(SDW) fluctuations
of relaxation type. 
 In the present paper  we show that this state 
transforms into SDW quantum liquid of resonance type
(quantum spin liquid) when the system passes to the SC
state. This means that the state below $T_{sc}(\delta)$ is
at the same time an ordered SC state with respect to
electronic degrees of freedom and a quantum spin liquid
state with respect to
spin degrees of freedom. This duality as we show is at the origin of the
observed by different neutron scattering groups specific
behaviour which includes the so called $40
meV$ resonance peak, incommensurability at lower $\omega$,
spin gap and other features.

In the case we consider, the fermionic 
subsystem is characterized by the spectrum

\begin{equation}
\epsilon_{{\bf k}} = \ -2t (\cos k_x + \cos k_y ) - 4t' \cos k_x
\cos k_y -...\ 
\label{1}
\end{equation}
It has a hyperbolic metrix in a proximity
of ETT:

\begin{equation}
\tilde{\epsilon}({\bf k}) =\epsilon_{{\bf k}}-\mu= -Z +a k_\alpha^2 -
b k_\beta^2, \hskip 0.4 cm Z \propto \delta_c-\delta.
\label{2}
\end{equation}
In (\ref{2}), 
$k_x$, $k_y$ are  distances from saddle-point (SP) wavevector
$\bf k$=${\bf k}_{SP}$=$(0,\pi)$,
$a$=$t-2t'$, $b$=$t+2t'$.
The important parameter $Z=\mu-\epsilon_{s}$
$\propto \delta_c-\delta$
($\epsilon_{s}=4t'/t$ is SP energy) measures the
energy and doping distance from ETT \cite{PRL1}
and determines a {\bf new energy scale} in the system.
In the SC state of the d-wave symmetry which
develops around ETT point,
the spectrum gets a gap
$\Omega_{\bf k}=\sqrt{\tilde\epsilon_{\bf k}^2+\Delta^2_{\bf k}}$,
 $\Delta_{\bf k}=\Delta(\cos k_x- \cos k_y)/2$,
$\Delta=\Delta_{(0,\pi)}$ which determines a {\bf second
low energy scale} in the system.
 SDW  fluctuations
in this SC state  in
   RPA
  approximation   are described by
$$
\chi({\bf q},\omega)=
\chi^0({\bf q}, \omega)/(1+V_{\bf q}\chi^0({\bf q},
 \omega)),$$
\begin{equation}
\chi^0({\bf q},\omega)=\frac{1}{N}\sum_{{\bf k},\pm}
M^\mp_{\bf qk}\frac{1-n^{F}(\pm\Omega_{{\bf k}})
-n^{F}(\Omega_{{\bf {q+k}}})}{{\Omega}_{{\bf q+k}
} \pm
{\Omega}_{{\bf {k}}}-\omega-i0^+}
\label{3}
\end{equation}
where $M^{\pm}_{\bf qk}=
1 \pm (\tilde\epsilon_{\bf k}\tilde\epsilon_{\bf k+q}
+\Delta_{\bf k}\Delta_{\bf k+q})/\Omega_{\bf k}\Omega_ {\bf k+q}$.
The interaction
entering in (\ref{3})
comes from the $J$-term in the $t-t'-J$ model: 
 $V_{\bf q}=J_{\bf q}=2J(\cos{k_x}+\cos{k_y})$.

As known, in the normal state ($\Delta=0$)
$\chi^0({\bf q}, 0)$ diverges as $Z \rightarrow 0$, 
$T \rightarrow 0$, ${\bf q
\rightarrow Q}=(\pi,\pi)$ so that SDW instability with
${\bf q
= Q}$ obligatory occurs in a vicinity of ETT point.
 The metallic state above keeps a memory
about this instability anomalously far in doping \cite{PRL1}.
The
parameter $\kappa^2=1-|J_{\bf Q}| \chi^0({\bf Q}, 0)$ which
describes a "proximity"  to SDW instability increases
anomalously slow with increasing a doping distance from
the instability. [This occurs only in the underdoped
regime $Z>0$ ($\delta<\delta_c$). The reason is analysed in \cite{Capecod}].
In the SC state
 $\chi^0({\bf Q},0)$ behaves as in 
 the normal state when $Z  \gg \Delta$
while the behaviour changes
strongly  when  $Z \leq \Delta$, see Fig.1a. In the latter case
 instead to diverge  $\chi^0$ strongly decreases
as $Z \rightarrow 0$. As a result,
 SDW instability around ETT point, $Z=0$, disappears 
but a memory about it (a SDW liquid state with small $\kappa^2$) 
 exists at larger $Z$.
 So far as for the cuprates 
$\Delta(\delta) \approx Const$  \cite{Timusk}
while 
$Z(\delta) \sim \delta_c-\delta$ by the definition, this means the
existence of two doping regimes   within the
underdoped regime, the first
close to optimal doping, $\delta^*<\delta<\delta_c$, for which
$\kappa^2$ (i.e. a disorder) increases with increasing
 doping and the second $\delta<\delta^*$ where it behaves in
the opposite way, see Fig.1b.
The point $\delta=\delta^*$ turns out to be a point of
a minimum disorder in this SDW liquid \cite{SDW}. A value of $\delta^*$
depends on $\Delta/t$ for fixed $t'/t$.

\begin{figure}
\vspace*{-1.5cm}
\hspace*{-1.cm}
\epsfig{%
file=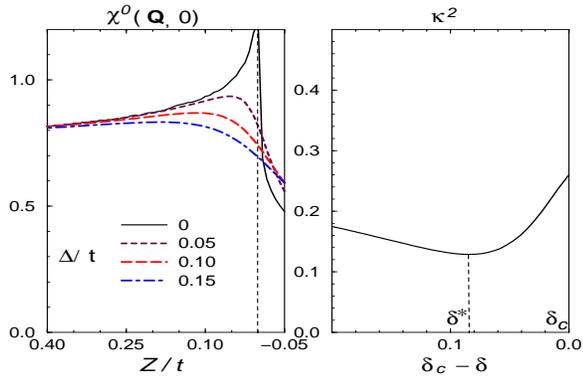,%
figure=fig1.ps,%
height=10cm,%
width=7.083cm,%
angle=-90,%
}
\\
\vspace*{-0.8cm}
\caption{Doping dependences of
$4J\chi^0({\bf Q},0)$    and $\kappa^2$  
 in the underdoped regime $Z>0$ ($\delta<\delta_c$).
$\chi^0$ is
presented as a function of $Z$,
$\kappa^2$ as a function of $\delta_c-\delta$ ($T=0$). 
Here and  later on $t'/t=-0.3$, 
$t/J=2$ [16]. $\delta_c=0.27$ for $t'/t=-0.3$.}
\label{f1}
\end{figure}

\begin{figure}
\vspace*{-1.5cm}
\hspace*{-0.cm}
\epsfig{%
file=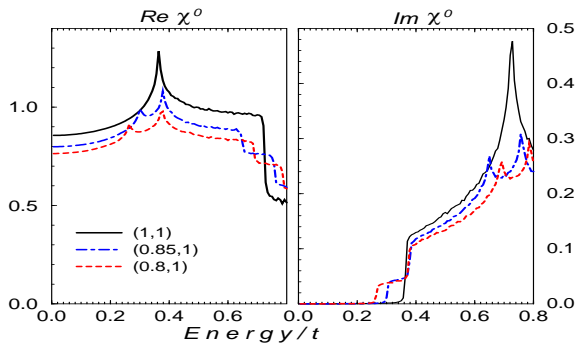,%
figure=fig2.ps,%
height=9.09 cm,%
width=6.44 cm,%
angle=-90,%
}
\\
\vspace*{-0.6cm}
\caption{ $J_{\bf q} Re\chi^0({\bf q},\omega)$  
and $JIm\chi^0({\bf q},\omega)$  for different
${\bf q}$ in the  direction $(q_x,1)\pi$.
$Z/t=0.2$, $\Delta/J=0.2$. $T=0$}
\label{f2}
\end{figure}

\begin{figure}
\vspace*{-1.5cm}
\hspace*{-1.cm}
\epsfig{%
file=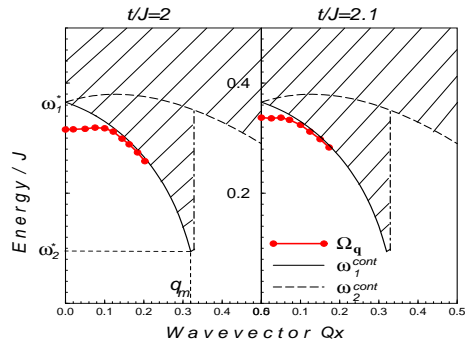,%
figure=fig3.ps,%
height=9.09 cm,%
width=6.44 cm,%
angle=-90,%
}
\\
\vspace*{-0.8cm}
\caption{Energies of the electron-hole continuum
$\omega_{1,2}^{cont}({\bf q})$ 
and of the  spin-exciton mode $\Omega_{\bf q}$
as  functions of wavevector for
 the direction $(1-Q_x,1)\pi$. $Z/t=0.2$, $\Delta/J=0.2$.}
\label{f3}
\end{figure}

\begin{figure}
\vspace*{-1.5cm}
\hspace*{-1.cm}
\epsfig{%
file=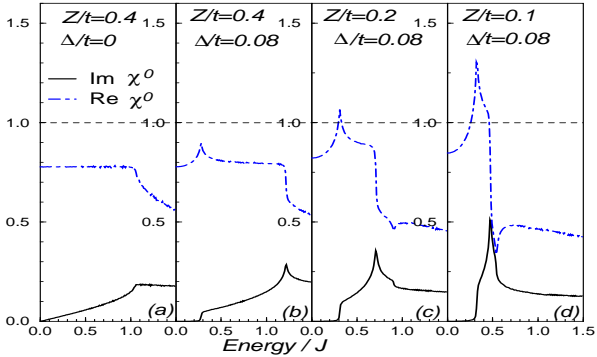,%
figure=fig4.ps,%
height=10.5 cm,%
width=7.437 cm,%
angle=-90,%
}
\\
\vspace*{-0.8cm}
\caption{Energy dependences of $Re\chi^0$ 
and $Im\chi^0$ for ${\bf q=Q}$
(in units of $1/4J$)
for the  normal state  (a), for the highly
underdoped
regime, $Z \gg \Delta$, (b) and for
 moderate  and close to optimal doping, $Z \sim \Delta$ or
$Z < \Delta$, ((c), (d).}
\label{f4}
\end{figure}

Let's study now a spin {\bf dynamics} in the SC state.
For this let's firstly analyse $\chi^0$ as a function
of $\omega$ and $\bf q$.  Most important features are summarized
in Fig.\ref{f2}-\ref{f4}.  
In Fig.\ref{f2} we show $\chi^0$ calculated
numerically  as a function of energy 
 for different wavevectors in a vicinity of ${\bf q=Q}$.

Analytical calculations performed simultaneously
show that for ${\bf q=Q}$,
$Re\chi^0$ diverges
logarithmically while  $Im\chi^0$ has
a jump at the energy \cite{ref}

\begin{equation}
 \omega=\omega^*_1=2\Delta_{} \sqrt{1-Z/|4t'|-(\Delta/8t')^2}
\label{5}
\end{equation}
 corresponding
to  opening of the gap in $Im\chi^0({\bf Q},\omega)$
while
for ${\bf q \neq Q}$ the  logarithmic singularity 
{\bf splits into two
logarithmic singularities}. The first one occurs  at
the energy corresponding to the low energy border in
 $Im\chi^0({\bf q},\omega)$, $\omega=\omega_1^{cont}({\bf q})$,
the second, $\omega=\omega_2^{cont}({\bf q})$,
lies in the continuum of electron-hole excitations.
[At both energies   $Im\chi^0({\bf q},\omega)$ jumps.]
These two  energies are important characteristics of  
the continuum, we
 plot them as functions of wavevector in Fig.\ref{f3}. 
One can see that 
the low energy border of the continuum
 behaves in a very nontrivial way. It
 is highly nonmonotonical so that for
$\omega < \omega_1^{cont}({\bf Q})=\omega_1^*$, spin-flip electron-hole excitations
exist in an extremely narrow $\bf q$ range.
The important wavevector is ${\bf q}_{m}$ \cite{qm}, 
see Fig.\ref{f3}.
Its absolute value in the direction
$(q_x,1)\pi$ (which as we will see later on corresponds
to maximum $Im\chi$ for given $\omega$)
and the corresponding energy of 
the continuum are given by
$$\pi q_{m} =2 arccos{[(|4t'|-Z)/2t}],$$
\begin{equation}
\omega_2^*=\omega_1^{cont}({\bf q}_m)=
\Delta (|4t'|-Z)/2t.
\label{7}
\end{equation}
As $Z \sim \delta_{c} -\delta$, both
$\omega_2^*$ and the deviation of $|{\bf q}_m|$ from $(\pi,\pi)$  
increase with increasing doping.
$\omega_2^*$ determines
a minimum gap in the spin excitations  for given direction
\cite{nodes}. Another outstanding point
is ${\bf q=Q}$, $\omega=\omega^*_1$: for  
$\omega>\omega^*_1$, the continuum occupies the whole $\bf q$
space. $\omega^*_1$ also increases with increasing doping, see eq. (\ref{5}). For fixed $Z$ (doping) both
 $\omega_{1,2}^{cont}({\bf q})$ are proportional to 
the SC gap $\Delta$ (when $|\Delta/8t'| \ll 1$).

Let's consider now features of total $\chi({\bf q},\omega)$,
i.e. a {\bf collective} spin dynamics.
 The divergence of $Re \chi^0$ for ${\bf q}$
in the vicinity
of $(\pi,\pi)$ means that   
there are  poles in
 $\chi({\bf q},\omega)$ at  energies
for which

\begin{equation} 
1-|J_{\bf q}|Re\chi^0({\bf q},
 \omega)=0.
\label{8}
\end{equation} 
 One of them obligatory lies below the
electron-hole continuum (at some $\omega=\Omega_{\bf q}$)
and therefore
$Im\chi^0({\bf q},
 \Omega_{\bf q})=0$. 
Spin susceptibility near this
pole is given by

\begin{equation}
Im\chi({\bf q},\omega)=(1/J_{\bf q}^2)[\partial
Re\chi^0({\bf q},
 \omega)/\partial\omega]^{-1}
\delta(\omega-\Omega_{\bf q}).
\label{9}
\end{equation}
This expression describes a {\it collective mode of  resonance
type - spin exciton}.  Details about it 
 depend on details of 
$\chi^0({\bf q}, \omega)$. One has to distinguish
two different situations: (i)  $Z \gg \Delta$ 
  (highly underdoped regime) and (ii)  $Z \sim \Delta$
or $Z < \Delta$
  (moderate and close to optimal doping).

In the case (i) the energy dependences of $Re\chi^0$ and 
$Im\chi^0$ are  close to those in the normal state (compare
Fig.\ref{f4}a,b).
The logarithmic singularity appearing in  $Re\chi^0$
is quite weak so that any 
extra factor leading to a finite damping
such as small amount of impurities, a temperature
effect etc. easily eliminates the singularity and
 the condition (\ref{8})
is not fulfilled. On the other hand, the jump in
 $Im\chi^0$ is  small, therefore due to the same damping
the gap in  $Im\chi^0$ is easily filled up.
As a result the {\bf spin dynamics in the
highly
underdoped regime is close to that
in the normal state}. This  is exactly what is observed 
in  $YBCO_{6.5}$ \cite{6.5}].

The situation 
changes in the case (ii) of relatively small $Z$.
  It is related to the fact that for $Z \sim 0$ (more
precisely for $Z=\Delta^2/8t'$) the type of  singularity at
$\omega=\omega^{cont}_1({\bf Q})$ changes from
logarithmic  to inverse square-root:  $Im\chi^0$, 
$Re\chi^0$  $\propto \sqrt{1/(\omega^{cont}_1-\omega})$.
As a consequence, in the case (ii) 
 the logarithmic behaviour still takes place but now only 
 in a narrow
 vicinity of $\omega=\omega^{cont}_1({\bf q})$ while at lower
energies  the inverse square-root behaviour in
 $Re\chi^0$ is preserved.
 This means that the $\omega$ dependence of
  $Re\chi^0$ is much stronger
for low energies than in the case (i)
while the gap in  $Im\chi^0$ is strongly marked
(see Fig.\ref{f4}c,d), therefore the
condition  (\ref{8}) for the pole is always fulfilled.
As a result, in this regime
the resonance mode exists  below
the continuum, the energy distance between this  mode and the
continuum for fixed $Z$ depends on $J/t$, see 
 Fig.\ref{f3}.
 The factor before $\delta$-function in (\ref{9})
i.e. the intensity of the resonance mode is 
high when the latter is far from the continuum, it
decreases with approaching the continuum, as clear from eq.(\ref{9})
and Fig.\ref{f2}a.

A picture of $\bf q$ dependence 
of the total susceptibility in this regime is qualitatively
different for different $\omega$ ranges.

1. For $\omega<\omega_2^*$ there is no magnetic
fluctuations neither of two-particle 
nor of collective nature.  This is a "spin gap" regime \cite{nodes}.
As we discussed above, $\omega_2^*$ i.e. the "spin gap"
increases with increasing doping that explains well the
observed by INS general tendency.

\begin{figure}
\vspace*{-3cm}
\hspace*{-3cm}
\epsfig{%
file=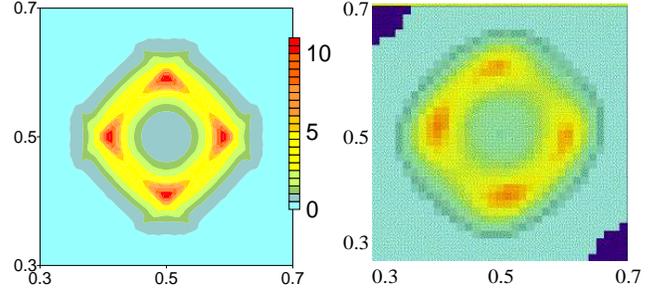,%
figure=fig5.ps,%
height=12cm,%
width=8.5cm,%
angle=0,%
angle=-90,%
}
\\
\vspace*{-1.5cm}
\caption{(color)  $\bf q$ dependences of $Im\chi({\bf q},\omega)$
for the energy regime 2
(a) theoretical  ($\omega/J=0.25$, other
parameters are as in Fig.3a),  (b) experimental
[7] ($YBCO_{6.6}$,  $\omega=25 meV$). The point $(0.5,0.5)$
corresponds to $(\pi,\pi)$.}
\label{f5}
\end{figure}

2. For $\omega>\omega_{2}^*$, rather strong incommensurate
spin fluctuations appear with a very
narrow $\bf q$-width. Their nature changes progressively
(with increasing $\omega$) from two-particle excitations to
the collective resonance  excitations; respectively
the intensity increases with increasing $\omega$. The ${\bf q}$ dependence of $Im\chi({\bf q},\omega)$
for this $\omega$ range is
shown in Fig.\ref{f5}a.
 It is strikingly similar to that observed recently experimentally
for $YBCO_{6.6}$ \cite{incommensurability},
see Fig.\ref{f5}b.

3. A maximum intensity of spin fluctuations occurs
at $E_r$=$\Omega_{\bf Q}$.
Due to  flatness of the dispersion  $\Omega_{{\bf q}}$,
the $\bf q$-width of $Im\chi({\bf q},\omega)$
 is much larger for $\omega \approx E_r$ than for energies
corresponding to the incommensurate
fluctuations that is in a good agreement with experiment 
\cite{incommensurability}.

4. For $\omega>\omega^*_1$, the spin fluctuations
has a two-particle nature, the $\bf q$ dependence is very broad.
There is still a maximum of intensity at the incommensurate
position corresponding to another pole  determined by
eq.(\ref{8}) which lies inside the continuum. This  is also in a good agreement with experiment
\cite{Arai}.

The obtained picture of the spin dynamics combining
a resonance mode plus two-particle continuum features
is typical for quantum spin liquid (QSL). Until now the latter
 was known as an attribute of low dimensional 
localized spin systems \cite{quantumliquid}. We have shown 
 that SC state in 2D electron system close
to ETT is another example of
QSL state.
 Is d-wave symmetry of SC gap  
necessary for this? The answer is no. For the isotropic s-wave
and for all other symmetries except  the extended s-wave 
(this
exception was already emphasized in \cite{Onufrieva}) there is always a
gap in $Im\chi^0({\bf Q},\omega)$, i.e.
the {\bf necessary} condition
for the existence of the resonance mode with $\bf q$
close to $\bf Q$ is fulfilled.
The difference with the d-wave symmetry (or with any other
symmetries for which the SC gap changes sign
 between different sectors of BZ,  d+s ...)
is that  due to the absence of singularity
 in $Re\chi^0$ \cite{cohfactor}
 the resonance mode   exists only when
 $J/t$ exceeds some threshold value.
 Another more significant  difference is that
in the case of the isotropic s-wave
symmetry the low energy border of the continuum 
 does not depend on wavevector,
  as a result the effect of  incommensurability
at low $\omega$ disappears.
On the other hand, two conditions, d-wave symmetry of SC gap and
a proximity to ETT quantum critical point
 (small
$Z$  or a proximity 
of Fermi level to SP) are {\bf sufficient} for the SC state
to be  a QSL.
Both conditions are known  to be fulfilled
for the hole-doped high-$T_c$ cuprates.

\begin{figure}
\vspace*{-1.5cm}
\hspace*{-1.cm}
\epsfig{%
file=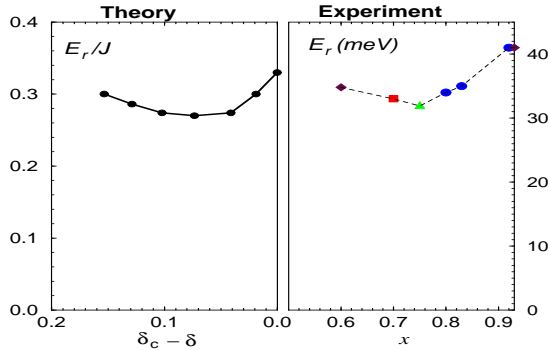,%
figure=fig6.ps,%
height=9.52 cm,%
width=6.745 cm,%
angle=-90,%
}
\\
\vspace*{-0.8cm}
\caption{Doping dependence of the resonance peak 
in the underdoped regime:  theoretical calculated with
$\Delta/J=0.2$, $t/J=2$  and  experimental for $YBCO_{6+x}$ 
[2-7].}
\label{f6}
\end{figure}

Coming back to the cuprates let's analyse 
  a doping dependence of the spin exciton energy. 
The latter is given by the condition  (\ref{8}). If it 
were fulfilled for low $\omega$,
the dispersion law  would be written
as $\Omega_{\bf q}$=$ \sqrt{\Delta^2_{r} +
a({\bf q-Q})^2}$
with $\Delta^2_{r} \propto \kappa^2$ (since
$Re\chi^0({\bf q},\omega) \propto \omega^2$ for low $\omega$)
that is a typical dispersion law for a quantum spin liquid.
For  realistic for the cuprates 
values of
 parameters, $t/J$$\sim$$2$, $\Delta/J$$\sim$$0.2$, the
condition (\ref{8})  is  fulfilled  for relatively high
$\omega$ for which roughly $\Delta_{r} \propto \kappa^2$.
 These simple arguments explain  the calculated
 doping dependence  $E_r(\delta) \equiv \Omega_{\bf Q}(\delta)$  which follows qualitatively
 the doping dependence of $\kappa^2$, compare   Fig.\ref{f6}a
 with  Fig.\ref{f1}b.
The experimental  dependence
 for the underdoped $YBCO_{6+x}$ is shown in 
Fig.\ref{f6}b where
 we summarized all reported by
different INS groups data for 
moderate  and close to optimal doping.
 The  theoretical
and experimental curves are strikingly similar.
 [Note when comparing that for the cuprates $J \approx 120meV$.]

Summarising, the study performed in the present 
paper has a significance
in two points.

1. We discover a new example of a quantum spin liquid
which is a 2D itinerant spin system in the SC state.
 As a quantum spin liquid in low dimensional localized
 spin systems
 it is characterized by the resonance
spin mode and by well separated two-particle
continuum.

2. The  picture obtained with using parameters
characteristic for the  high-$T_c$ cuprates
allows to shed light on many observed 
in the underdoped regime features such as the very fact
of incommensurability at low $\omega$,
 the absolute value of the incommensurability wavevector
and the details of $\bf q$
dependence of $Im\chi$ in different directions in BZ,
 the absolute value of the resonance energy $E_r$
 for optimal doping and its doping dependence, 
the interrelation between the $q$-width of $Im\chi$ in the case of
commensurate and incommensurate fluctuations etc.
The agreement with experiment 
 is quite impressive regarding
that we did not
use any external hypothesis or adjustable parameters 
\cite{parameters}.

Note that we did not consider the case of 
 $|t'/t| \rightarrow 0$
which corresponds to another class of universality
as we emphasized in \cite{PRL1}. This case
 (presumably corresponding to
LSCO) for which an incommensurability
exists both below and above $T_{sc}$ will be analysed elsewhere.

We would like to thank the
Aspen Center for Physics for their hospitality
during the High-$T_c$ Summer Workshop 1998
where this work was initiated.


\begin{thebibliography} {10}


\bibitem{ARPES} H.Ding et al,
Nature (London), {\bf 382}, 51 (1996)



\bibitem{6.92}J.Rossat-Mignod et al, Physica C,
{\bf 185-189}, 86 (1991)

\bibitem{6.7}H.F. Fong et al, Phys.Rev.Lett. {\bf 78}, 713 (1997)

\bibitem{6.75}M.Sato et al, J.Phys.Soc.Jpn.  {\bf 62}, 263 (1993)

\bibitem{6.8}P. Bourges et al, Europhys.Lett. {\bf 38}, 313 (1997)

\bibitem{6.83}J.Rossat-Mignod et al, Physica B
{\bf 199-200}, 281 (1994)



\bibitem{incommensurability}H.A. Mook et al. Nature,
{\bf 395}, 580 (1998)

\bibitem{BSCO}H.F. Fong et al, Nature, to appear

\bibitem{Arai}M.Arai et al, to be published


\bibitem{PRL1}F.Onufrieva et al,  Phys.Rev.Lett. 
{\bf 82}, N 11 (1999)


\bibitem{PRL2}F.Onufrieva,  P.Pfeuty,  Phys.Rev.Lett.
{\bf 82},  N 13  (1999)

\bibitem{Onufrieva96}F. Onufrieva et al,  Phys.Rev.B
{\bf 54}, 12464 (1996)


\bibitem{Capecod}F.Onufrieva et al,J.Phys.Chem.Sol.,
{\bf 59}, 1853 (1998)


\bibitem{Timusk}T.Timusk et al, Rep.Prog.Phys,
to appear

\bibitem{SDW}For very large $J/t$ the SC state becomes 
unstable against SDW ordering in the vicinity
of $\delta=\delta^*$. The details about mixed SC+SDW
ordered state see elsewhere.



\bibitem{parameters} $t'/t$ is chosen to fit experimentally
observed in YBCO and BSCO form of Fermi surface, $t/J$ to correspond to
 theoretical estimations for the $t-J$ model,
$\Delta/J$ to correspond to ARPES and tunneling data.



\bibitem{ref}This was already discussed in \cite{Onufrieva,Mazin}



\bibitem{Onufrieva}F.Onufrieva, Physica C, {\bf 251}, 348 (1995)

\bibitem{Mazin} I.Mazin et al, Phys.Rev.Lett. {\bf 75}, 4134 (1995)


\bibitem{qm}$\bf q_m$ is determined by the
doping distance from the second quantum critical point
in the system at $Z=Z'_{c}=-4t'$ \cite{Capecod}. 
The details will be discussed
elsewhere.



\bibitem{nodes}For
 $(1,1)$ direction this gap  equals  zero due to  existence of
nodes in one electron spectrum. However the intensity is extremely
small and hardly can be detectable by INS.


\bibitem{6.5} P. Bourges et al,  Phys.Rev.B
{\bf 56}, R11439 (1997)




\bibitem{quantumliquid}S. Ma et al, Phys.Rev.Lett. {\bf 69}, 3571 (1992);
S. Chakravarty et al, Phys.Rev.B {\bf 39}, 2344 (1989)



\bibitem{cohfactor}The amplitude of the
logarithmic singularity  tends to zero due to vanishing of the coherence factor in (\ref{3})















\end{thebibliography}
\end{document}